\documentstyle[preprint,aps]{revtex}

\newcommand{\hell}{${}_{\Lambda}^3$H}

\begin{document}
\draft
\title{
On the Impossibility to Measure the Total 
Neutron- and Proton Induced Nonmesonic Decays for \hell
}
\author{J. Golak and H. Wita\l a}
\address{Institute of Physics, Jagellonian University, 
PL-30059 Cracow, Poland}
\author{K. Miyagawa}
\address{Department of Applied Physics,
                     Okayama University of Science,
                     Ridai-cho Okayama 700, Japan
}
\author{H. Kamada\footnote{present address: 
Institut f\"ur Strahlen- und Kernphysik der Universit\"at Bonn,
Nussallee 14-16, D-53115 Bonn,
    Germany   } and W. Gl\"ockle }
\address{Institut f\"ur Theoretische Physik II, 
Ruhr-Universit\"at Bochum, D-44780 Bochum, Germany}

\date{\today}
\maketitle

\begin{abstract}
Based on realistic calculations for the nonmesonic decay rate of 
\hell  we demonstrate, that in principle it is not possible 
to measure the total n- and p- induced decay rates and as a 
consequence $\Gamma _n $/$\Gamma _p $ for that lightest  hypernucleus.
For the nonmesonic decay process  
the calculations are performed with modern YN forces based on various meson 
exchanges and taking the final state interaction among the three 
nucleons fully into account. 
Our findings have consequences also for the interpretation of experimental 
$\Gamma_n$
/$\Gamma _p $ ratios for heavier hypernuclei where severe discrepancies exist
to theoretical $\Gamma_n$/$\Gamma_p$ ratios.
\end{abstract}


\narrowtext

\section{Introduction}
There is a longstanding discrepancy between the theoretical 
ratio of the total neutron induced nonmesonic decay rate 
$\Gamma _n $ to $\Gamma _p $, the total decay rate for the 
proton-induced nonmesonic decay rate of various hypernuclei to
experimental data\cite{1}. The experimental values are typically around
1 except for the very light hypernucleus $^4_\Lambda$He 
\cite{Outa,Zeps},
while 
theoretical evaluations  lead to 0.05 - 0.2.  The experimental value for $\Gamma 
_n $ is estimated either from neutron measurements and/or deduced from the measured values of the total nonmesonic 
decay rate $\Gamma_{nm} $ and of $\Gamma_p$ as 
\begin{eqnarray}
\Gamma_n \equiv \Gamma_{nm} - \Gamma_p 
\label{eq0}
\end{eqnarray}
Apparently this relation can not be strictly true 
due to interferences. 
The quantity $\Gamma_p$ is determined experimentally from measuring 
single proton spectra and assuming that those protons are generated 
by the p-induced decay.
Again this can not be strictly true since the n-induced decay leaves 
behind spectator proton(s) and final state interactions  can 
carry momentum from neutrons to protons. We shall shed light in this article
on those critical issues. On the theoretical side one faces the nuclear many
body problem. Rigorous solutions based on realistic modern baryon-baryon
forces are not in sight. Therefore shell model pictures supplemented 
by Jastrow type two-body correlations are typically being used and final 
state interactions are established by optical potentials. 
It appears difficult to estimate quantitatively the uncertainty of 
the theoretical predictions. In such a situation 
a view on very light systems is of increasing interest. In the 3-baryon 
system bound and scattering states can be rigorously gained based on 
modern realistic baryon-baryon forces \cite{3}.
Therefore  uncertainties about the quality of the hypernucleus 
wavefunction and final 
state interactions are absent. In the four-body system first rigorous 
solutions for bound states ($^4_\Lambda$H and $^4_\Lambda$He) already 
appeared \cite{4}. 
The mesonic and nonmesonic decays of \hell have been calculated 
\cite{3,5} but there are only few data 
to compare with.  Some mesonic decay rates for \hell    
for which data are 
available agree rather well with that theory. 
Though there are state of the art calculations no data are available
 for the very small
nonmesonic  decay rates of \hell. 
We would like to use in this article that theoretical insight 
to throw light on the questionable issues mentioned above. 
In \cite{3} we found that the nonmesonic decays of \hell  
leading to a final 
deuteron and a neutron are suppressed by about a factor 10 
with respect to the full breakup processes.
Therefore we shall neglect those two-body fragmentation  decay channels
of \hell  in the following - except for pointing out that 
there a separation of n- and p- induced decays is clearly impossible.
This is already evident from the fact that 
$\Gamma_n ^{n+d} + \Gamma_p ^{n+d} $= 0.39$\times$10$^7$s$^{-1}$,
whereas the total 
n+d decay rate $\Gamma ^{n+d}$= 0.66$\times$10$^7$s$^{-1}$.
Clearly there is a strong interference between the n- and p-induced 
decays.

The exclusive differential   n+n+p decay rate has the form \cite{3} 
\begin{eqnarray}
d\Gamma ^{ n+ n+ p } & =& { 1 \over 2 } 
\sum _{m , m_1 , m_2, m_3 } \vert \langle 
\Psi ^{(-)} _{ \vec p \vec q m_1 m_2 m_3}
\vert \hat O \vert \Psi_{^3 _\Lambda H, m} \rangle \vert ^2 
2 \pi d \hat k_1 d \hat k _2 d E_1
\cr 
& \times & { { M_N ^2 k _1 ^2 k_2 ^2 } 
\over 
{ \vert k_1  (
2 k_2   + \vec {k_1} \cdot \hat {k_2} ) \vert } 
 }
\label{eq1}
\end{eqnarray}
Here
$\hat {k_1} $ and $\hat  {k_2}$ denote the  directions of two detected nucleons
(see \cite{3} for further information).
We have shown in \cite{3} that there are regions in phase-space which are
 populated by n- and p- induced decays and therefore an experimental 
separation for those contributions is impossible.
But there are also regions in phase-space which are rather cleanly  populated 
by either n- or p- induced processes, but not both. 
Therefore one has to be satisfied with certain fractions of $\Gamma_n$ and 
$\Gamma_p$, defined by integrations over certain subregions of the 
total phase-space. 
In this manner one can measure n- and p- induced process separately.
The fact that certain parts of the phase-space are populated by both processes
coherently makes it obvious that by no means one will be able to access
experimentally $\Gamma_n$ and $\Gamma_p$ separately.
Nevertheless we would like to demonstrate this explicitely in the approach 
to $\Gamma_p$ which is being used for heavier hypernuclei \cite{1}.
There one investigates the semiexclusive  decay process in which only one proton is detected. Therefore we shall study in this article 
the single differential decay rate
$d\Gamma $/$dE_p$ and in addition also $d\Gamma $/$dE_n$ and investigate whether they 
can be separated into n- and p- induced contributions and whether certain 
energy ranges are dominated by one or the other process.

Our results are based on rigorous solutions of the Faddeev equations for 
\hell  and the 3N final scattering states. 
We use the YN Nijmegen potential \cite{6} which includes $\Lambda$-$\Sigma$ 
conversion. It turned out that this potential produces the experimental 
\hell  binding energy without further adjustment\cite{7}. 
For the NN forces we used the Nijmegen' 93  potential \cite{8}.
We expect no dependence on the choice among the most modern NN potentials.
For the hypertriton this 
has been verified. 
The importance of the final state interaction is demonstrated by also 
presenting results where the 3N scattering state in the nuclear 
matrixelement occurring  in Eq. (\ref{eq1}) is replaced by 3N plane wave states. 
This extreme approximation will, like in \cite{3}, be  denoted 
by symmetrized plane wave impulse approximation (PWIAS), whereas the 
calculation with final state interaction will be called "FULL". 
In Fig. 1 we show $d\Gamma $/$d E_n$, $d \Gamma_n$/$d E_n$ and 
$d \Gamma_p$/$d E_n$ in PWIAS. The quantity $d \Gamma$ / $d E_n$ has 
two peaks, one  at very low neutron energies and one close to the maximal 
possible neutron energy. The peak at the higher energy is fed by the n- 
and p- induced processes as is obvious from the corresponding peaks in
$ d \Gamma _n $/$d E_n$ and $d \Gamma_p $/$d E_n$.
Clearly in both processes a high energetic neutron 
is produced. Surprisingly for us $d\Gamma_n $/$d E_n$ + $d\Gamma_p$/$d E_n$
sum up to $d\Gamma$/$d E_n$ with an error smaller than 5 \%.
The interference terms are therefore numerically very small. 
For very small neutron energies $d\Gamma _n $/$d E_n$ dies out, 
since the n- induced process creates mostly
 high energetic neutrons. The p- induced 
process, however, $d \Gamma_p$/$d E_n$, exhibits a strong peak at very low
neutron energies, which is caused by the (spectator) momentum distribution
of the neutron in \hell. Clearly a measurement of the decay rate
$d \Gamma $/$d E_n$
as a function of the neutron energy will not allow to separate the 
n- and p- induced processes - except at very low neutron energies, where the
energy distribution of the neutrons, however, is not determined by the 
$\Lambda$-decay process. That picture does not change qualitatively 
if one turns on the final state interaction as can be seen in Fig 2. 
Quantitatively, however, the rates are quite different. We can see 
a reduction factor of about 2 and the neglection of FSI would be 
disastrous 
in a quantitative analysis of data.
Now the sum $ d\Gamma _n$/$d E_n$ + $d \Gamma_p $/$d E_n $ equals  $ d \Gamma $/$d E_n$
only within about  12 \%. 

The situation for a separation of n- and p- induced processes appears somewhat 
more favourable 
if one regards the single particle decay rates as a function of 
the proton energy. Our results are shown in Fig. 3 for PWIAS and Fig. 4 
for the "FULL" calculation. For large proton energies nearly all protons 
result from the p- induced process : $d \Gamma $/$d E_p \approx d \Gamma_p$/
$d E_p$ in case of PWIAS. The quantity 
$d \Gamma_n$/$d E_p$ can not produce high energetic 
protons except due to FSI and this is indeed visible by comparing 
Figs. 3 and 4. 
$d \Gamma_n $/$d E_p$ exhibits, however, the very low energetic proton 
peak from the spectator proton in \hell.
Also note  again the reduction factor of about 2 caused by FSI. 

Let us now quantify the question, whether integrated proton distributions
can provide a good estimate for $\Gamma_p$. 
Clearly the very low energetic peak should be excluded and one has to 
start integrating $d \Gamma $/$d E_p$ from the highest possible 
proton energy $E_p ^{max}$, 
downwards. Thus we compare the integrals 
\begin{eqnarray}
\Gamma (E_p ) \equiv \int _{E_p }^{E_p^{max}} d E_p' { {d \Gamma
}\over {d E_p ' }} 
\label{eq4}
\end{eqnarray}
\begin{eqnarray}
\Gamma _p (E_p ) \equiv \int _ {E_p }^{E_p^{max}} d E_p ' { {d \Gamma 
_p }\over {d E_p ' }} 
\label{eq5}
\end{eqnarray}
and 
\begin{eqnarray}
\Gamma _n (E_p ) \equiv \int _ {E_p }^{E_p^{max}} d E_p ' { {d \Gamma
_n }\over {d E_p ' }}
\label{eq6}
\end{eqnarray}
as functions of $E_p$. The results are displayed in Figs. 5 and 6 for 
PWIAS and FULL. 
We see that in the case of PWIAS down  to about $E_p \approx $ 50MeV 
the two curves $\Gamma (E_p) $ and $\Gamma _p (E_p) $ are close to 
each other within less than  5\% and only then start to deviate strongly.
While $\Gamma _p (E_p)$ flattens out and approaches $\Gamma_p$=$\Gamma
(E_p=0)$, $\Gamma (E_p )$ receives  contributions from the 
n-induced process. The situation is not so favourable,
however, for the case FULL.
Around $E_p$=60 MeV the relative deviation $| \Gamma _p (E_p) - \Gamma
(E_p) | $/$ \Gamma_p(E_p)$ is about 10 \% and increase to about 20 \%
around $E_p$=15MeV. Below that the deviation increases up to 30 \%.
Note  also 
 the 
relative factor of about 2 between PWIAS and FULL.
We have to conclude that an estimate for $\Gamma_p$ from $
d\Gamma$/$dE_p$ is only possible within an error of about 30 \%. 
If one is satisfied with a fraction of $\Gamma_p$ the error can be
reduced to about 10 \%.

A well defined  manner 
to receive information on the p- and n-induced decays separately is
  to use the differential decay rate of Eq.(\ref{eq1})
as described in \cite{3}. There are certain regions in phase space which 
are populated only by the p- induced decay and others which
are populated 
only by the n-induced decay. In this manner one does not get the total 
$\Gamma_p $ or $\Gamma _n$, but at least well defined fractions thereof.

Now we would like to address the question, whether $\Gamma_n$ can be
found via Eq. (\ref{eq0}) in case of \hell. This is a pure theoretical
issue since, as we just demonstrated, $\Gamma_p$ can not be measured for
\hell.
Surprisingly enough Eq. (\ref{eq0}) is valid. 
As seen  from Table V 
in \cite{3} we have 
\begin{eqnarray}
\Gamma _n^{FULL} = 0.17 \times 10 ^8,  \Gamma_p ^{FULL}= 0.39 \times 10 ^8,
\Gamma _n^{FULL} + \Gamma _p ^{FULL} = 0.56 \times 10 ^8
\end{eqnarray} 
That sum has to be compared with 
$\Gamma ^{FULL}$= 0.57 $\times $10$^8$, which treats the full process
correctly as a coherent sum of the n- and p-induced decays.
 These numerical results validate Eq.(\ref{eq0}) in the 
case of \hell. 

Finally we note that our theoretical result for the  ratio of the total 
n- and p-induced decay rates in case of \hell 
is $\Gamma _n $/$\Gamma_p$ = 0.44.

Since $\Gamma_p $ in the case of \hell can not be measured, 
it appears advisable to concentrate directly on $d \Gamma$/$d E_p$
and $d \Gamma$/$d E_n$ and compare those distributions to theory. 
This is an alternative to the above mentioned exclusive processes.
While measurements of the nonmesonic decay of \hell  appear to be 
far away, data for the four-body hypernuclei already exist
\cite{Outa,Zeps} and 
theoretical predictions can be expected to come up in the near future.
This will then allow interesting tests of the nonmesonic decay 
matrixelements, which will be 
 based on realistic four-body wavefunctions and various 
meson-exchange operators \cite{3,10}, which drive the nonmesonic decay process.

\acknowledgments

This work has been  supported by the Deutsche Forschungsgemeinschaft 
(H.W. and H.K.) and by the Polish Comittee for Scientific Recearch
(J.G.)(grant No. 2P03B03914). 
The calculations have been performed on the CRAY T90
of the John von Neumann Institute for Computing, J\"ulich,
Germany.


\begin{figure}
\input{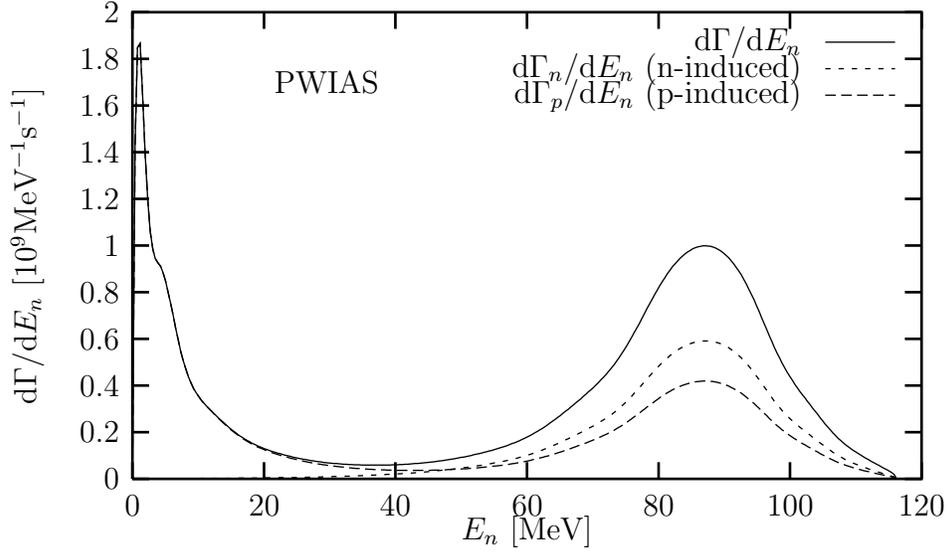}
\bigskip

\caption{
The single neutron decay rates $d \Gamma $/$d E_n$, $d \Gamma_n $/$dE_n$ and 
$d \Gamma_p $/$d E_n$ in PWIAS as a function of the neutron energy $E_n$.
A separation in n- and p-induced processes is not possible. The peak at very 
low $E_n$'s shows directly the momentum distribution of the neutron in 
\hell.
}
\label{fig1}
\end{figure}

\begin{figure}
\input{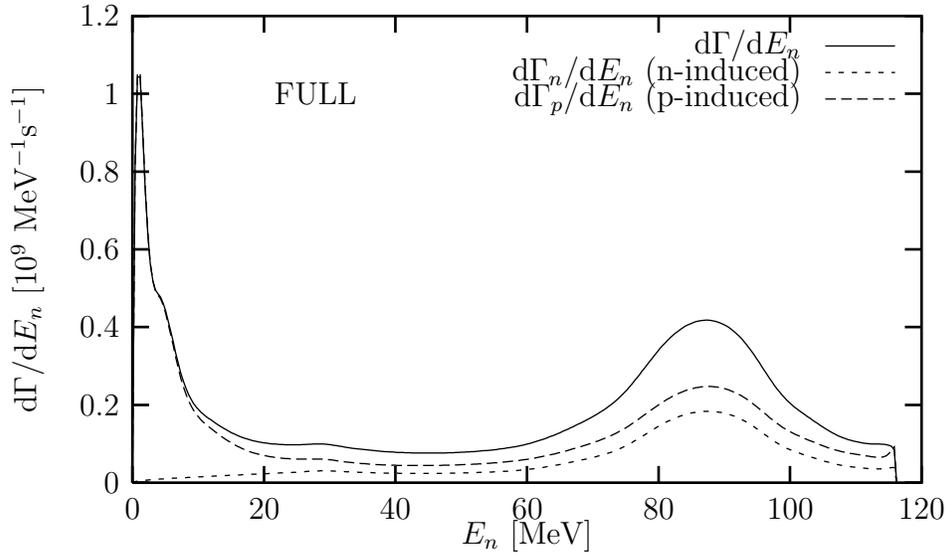}
\bigskip

\caption{
The same as in Fig. 1 for the FULL calculation. The peak at very low $E_n$'s
is now also influenced by  final state interactions.
}
\label{fig2}
\end{figure}

\begin{figure}
\input{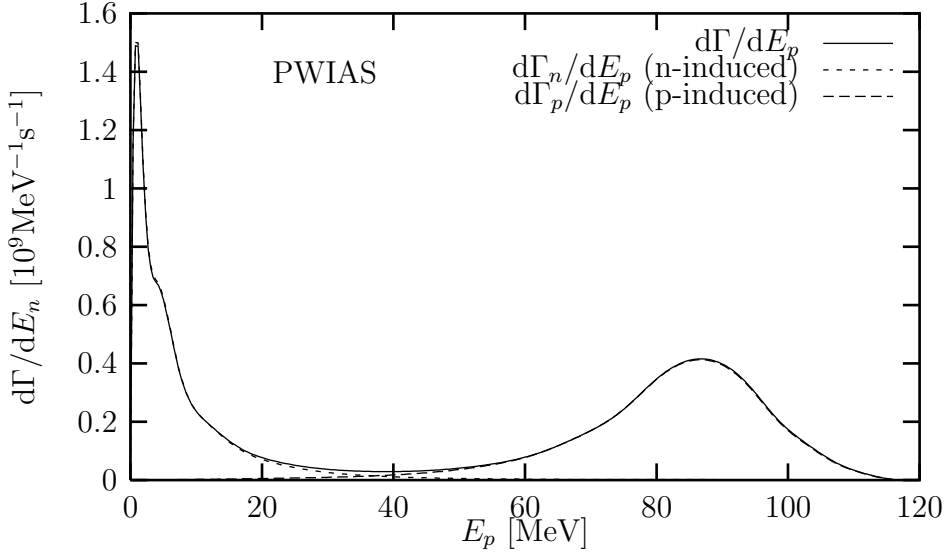}

\bigskip

\caption{
The single proton decay rates $d \Gamma $/$d E_p $, $d \Gamma_n$/$d E_p$ and 
$d \Gamma_p$/$d E_p$ in PWIAS as a function of the proton energy $E_p$. 
Now a separation in n- and p-induced processes would be possible for 
$E_p$ larger than about 50 MeV. The peak at very low $E_p$'s shows directly 
the momentum distribution of the proton in \hell.
}
\label{fig3}
\end{figure}
\begin{figure}
\input{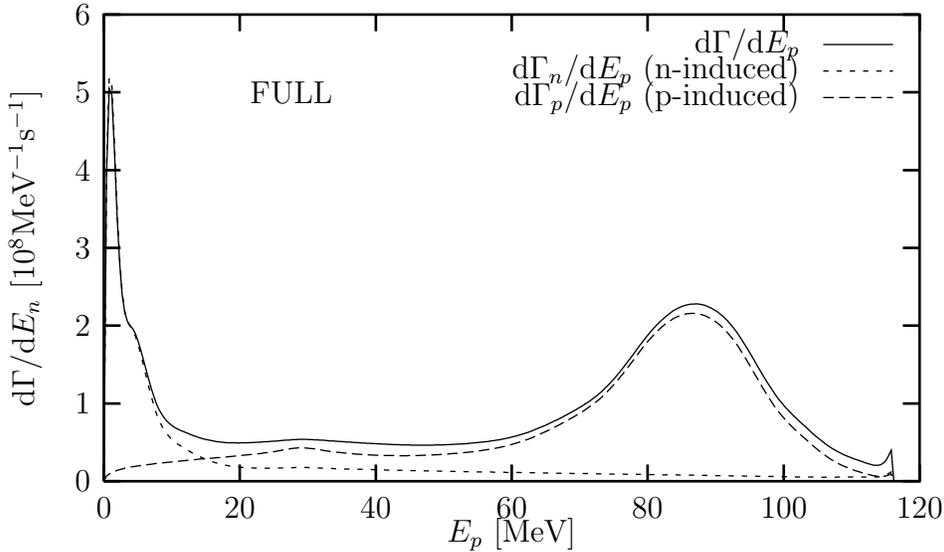}

\bigskip

\caption{
The same as in Fig. 3 for the FULL calculation. The final state interaction 
causes now small contributions of high energetic protons resulting from 
the n-induced decay. Also the peak at very low $E_p$'s is now influenced by 
final state interactions.
}
\label{fig4}
\end{figure}

\begin{figure}
\input{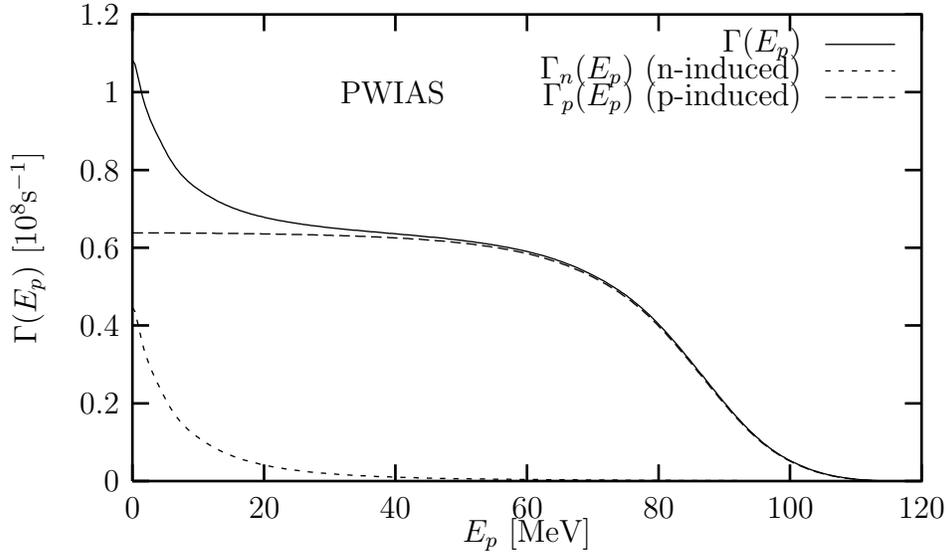}

\bigskip

\caption{
The integrated single proton decay rates according to Eqs.(\ref{eq4})-(\ref{eq6})
for PWIAS. For $E_p \ge $50MeV $\Gamma_p (E_p) \approx  \Gamma (E_p)$.
}
\label{fig5}
\end{figure}

\begin{figure}
\input{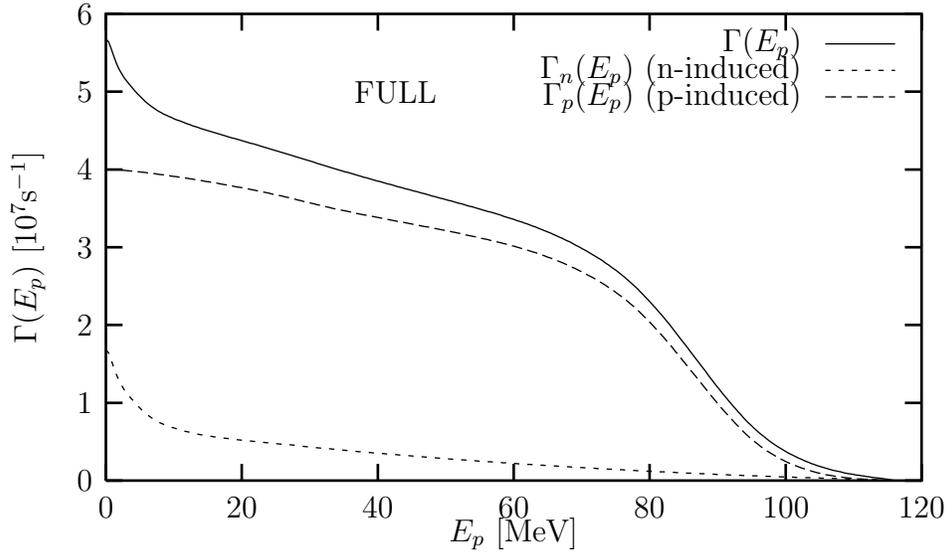}

\bigskip

\caption{
The same as in Fig. 5 for FULL. Now the influence of the n-induced decay does 
not allow to estimate $\Gamma_p(E_p) $ by $\Gamma (E_p)$.
}
\label{fig6}
\end{figure}


\begin{references}

\bibitem{1}
A. Ramos, A. Parre\'no, C. Bennhold, E. Oset, L.L. Salcedo, M.J.
Vicente-Vacas, Nucl. Phys. {\bf A 639}, 307c (1998). 

\bibitem{3}
J. Golak, K. Miyagawa, H. Kamada, H. Wita\l a, W. Gl\"ockle, A Parre\'no, 
A. Ramos, C. Bennhold, Phys. Rev. {\bf C 55}, 2196 (1997);
erratum, Phys. Rev. {\bf C 56},2982 (1997).
\bibitem{4}
E. Hiyama, M. Kamimura, T. Motoba, T. Yamada, W. Yamamoto, Nucl. Phys.
{\bf A 639 }, 169c (1998); E. Hiyama, in "Innovative Computational
Methods in Nuclear Many Body Problems", eds.: H. Horiuchi {\it et al.},
Osaka, 1997, World Scientific 1998, page 128.
\bibitem{5}
H. Kamada, J. Golak, K. Miyagawa, H. Wita\l a, W. Gl\"ockle, Phys. Rev. {\bf 
C 57},1595  (1998); W. Gl\"ockle, K. Miyagawa, H. Kamada, J. Golak, 
H. Wita\l a, Nucl. Phys. {\bf A 639}, 297c (1998).
\bibitem{6}
P.M.M. Maessen, Th. A. Rijken, J.J. de Swart, Phys. Rev. {\bf C 40}, 2226 (1989).
\bibitem{7}
K. Miyagawa, H. Kamada, W. Gl\"ockle, V. Stoks, Phys. Rev. {\bf C 51},2905 (1995).
\bibitem{Outa}
H. Outa, {\it et al.}, Nucl. Phys. {\bf A 639}, 251c (1998).
\bibitem{Zeps}
V.J. Zeps, Nucl. Phys. {\bf A 639}, 261c (1998).
\bibitem{8}
 V.G.J. Stoks, R.A.M. Klomp, C.P.F. Terheggen, J.J. de Swart,
   Phys. Rev. {\bf C49},  2950 (1994).
\bibitem{10}
A. Parre\'no, A. Ramos, C. Bennhold, Phys. Rev. {\bf C 56}, 339 (1997).

\end{references}
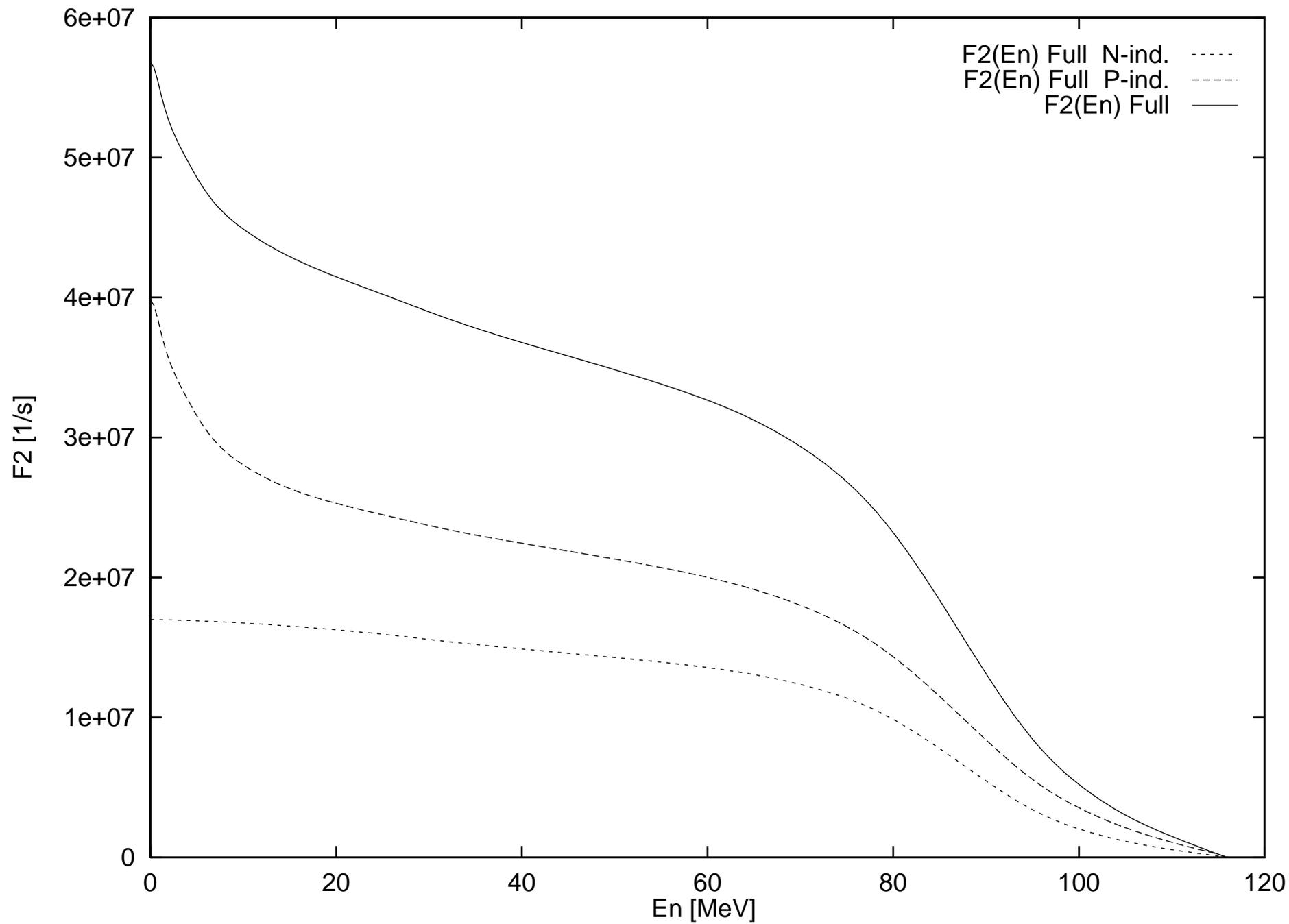
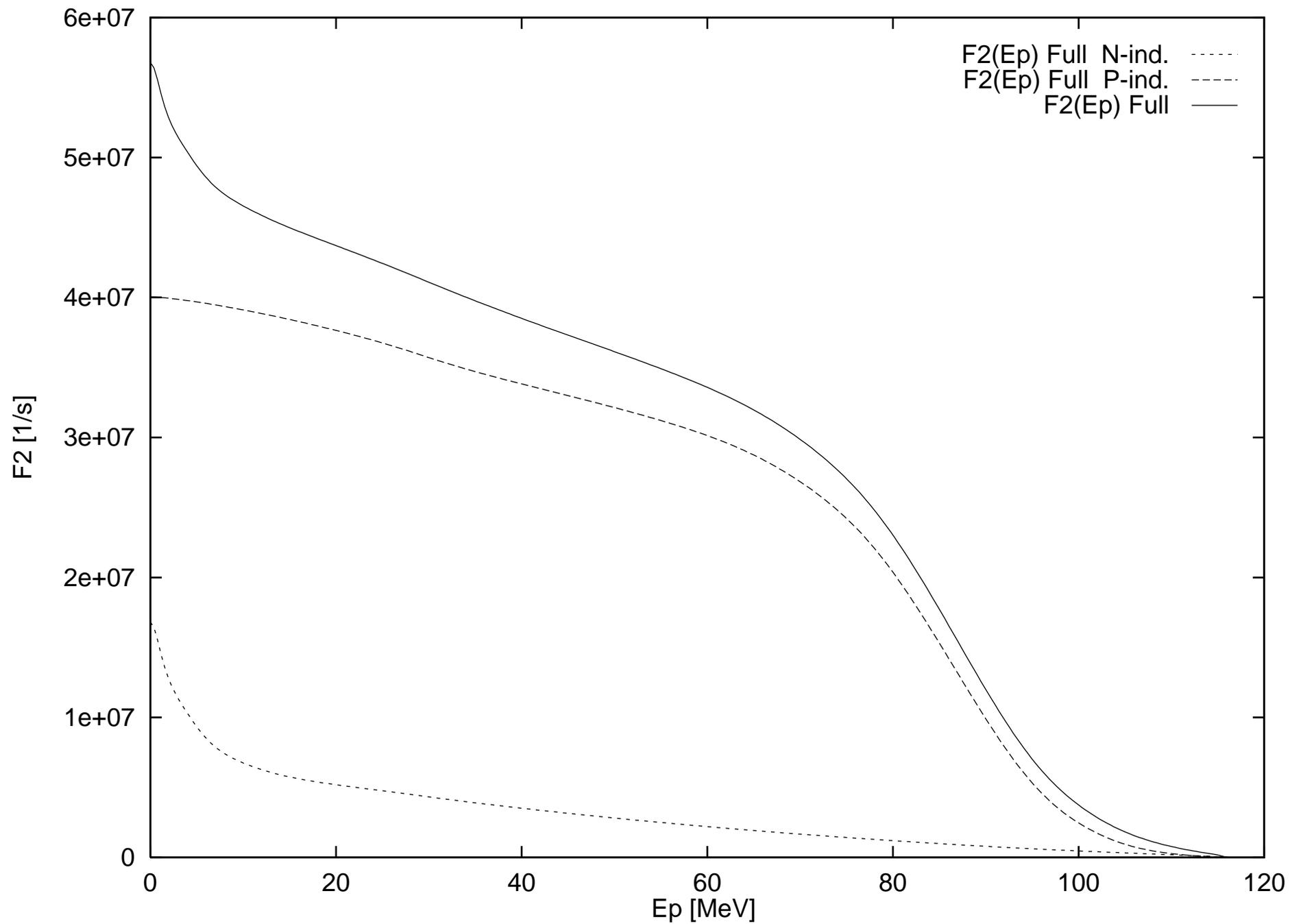
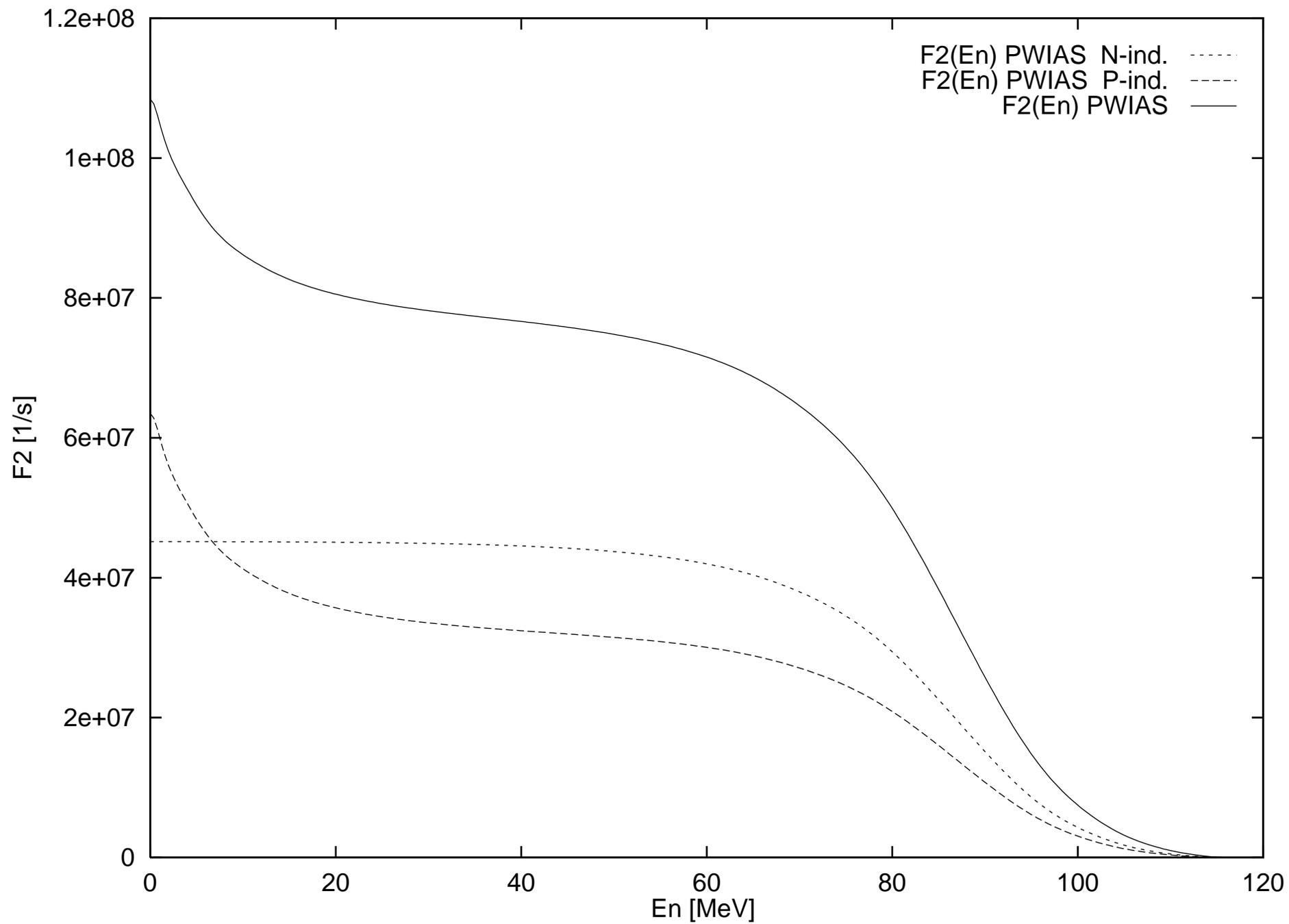
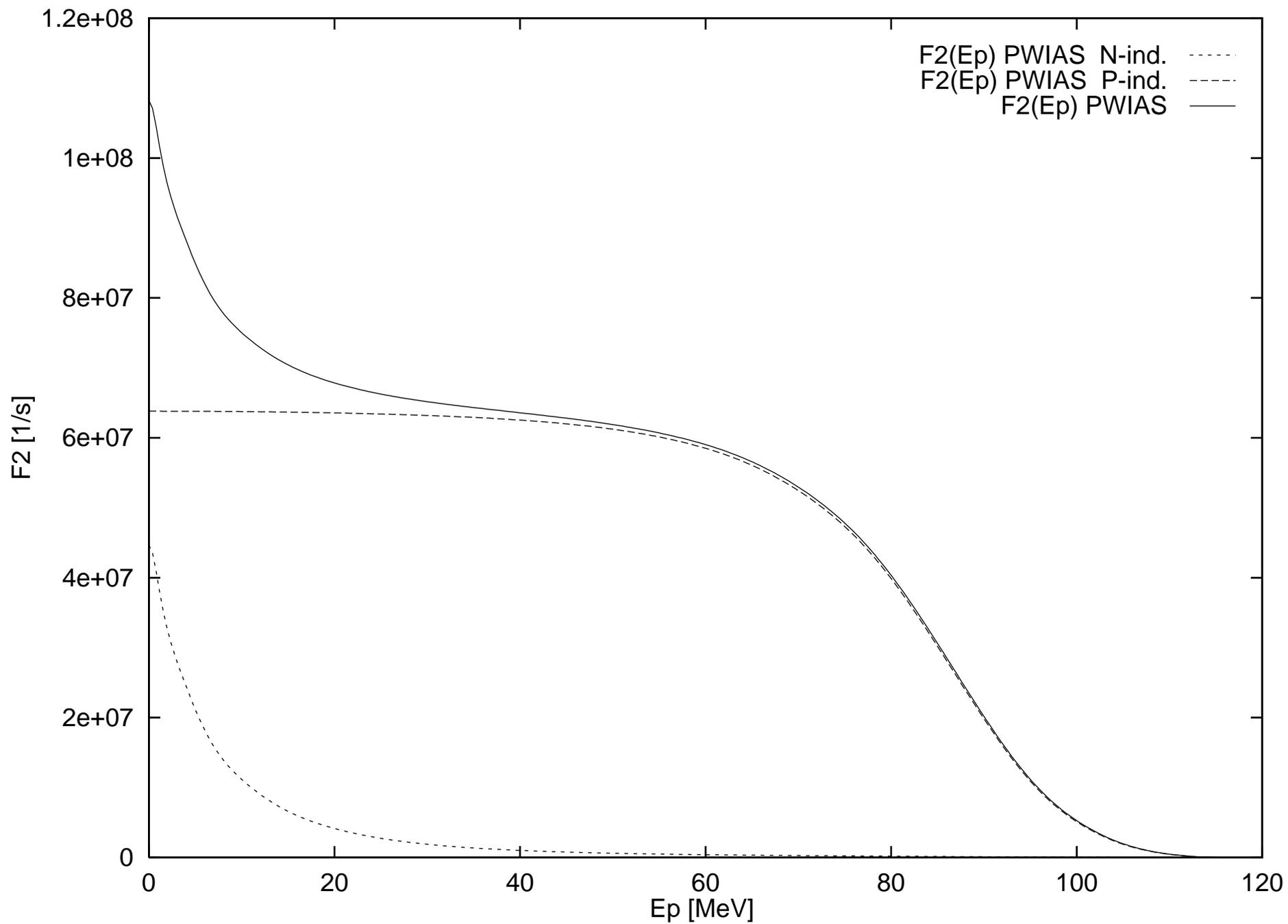
\end{document}